\documentclass[prb,twocolumn,superscriptaddress,showpacs, nofootinbib]{revtex4-2}
\usepackage{graphicx}
\usepackage{float}
\usepackage{dcolumn}
\usepackage{float}
\usepackage{bm}
\usepackage{mathrsfs}
\usepackage{ulem}
\usepackage[per-mode = fraction]{siunitx}
\usepackage{hyperref}
\hypersetup{
    colorlinks=true,
    linkcolor=blue,
    filecolor=magenta,      
    urlcolor=cyan,
    citecolor=magenta,
}
\usepackage{xcolor}
\usepackage{soul}
\usepackage{amsthm,amssymb}
\usepackage{mathtools}
\usepackage{comment}
\usepackage{physics}

\makeatother

\begin{document}
\title{Anomalous Seebeck effect and counter-propagating ballistic currents in graphene}
\author{A. V. Kavokin}
\affiliation{Westlake University, Hangzhou 310024, Zhejiang Province, China}
\author{S. V. Kavokina}
\affiliation{Westlake University, Hangzhou 310024, Zhejiang Province, China}
\author{A. A. Varlamov}
\affiliation{CNR-SPIN, Via del Fosso del Cavaliere, 100, 00133 Rome, Italy }
\author{Yuriy Yerin}
\affiliation{CNR-SPIN, Via del Fosso del Cavaliere, 100, 00133 Rome, Italy }
\date{\today }
\begin{abstract}
The Seebeck effect consists in the induction of a voltage drop due to the temperature difference in a conductor. In the middle of XIXth century, Lord Kelvin has proposed a relation between the Seebeck coefficient and the derivative of the chemical potential over temperature in the broken circuit regime. This relation appears to be nearly universal as it equally well applies to metals, semimetals and semiconductors. We show that it may fail, however, in graphene, due to the non-locality effects in the ballistic electronic transport regime. The correction to the Kelvin's formula emerges due to the coexistence of counter-propagating non-dissipative currents of cold and hot electrons. The external magnetic field normal to the graphene sample allows separating hot and cold currents in real space. The developed formalism may help interpreting the recent experimental data on ballistic edge currents in graphene bi-layers in the quantum Hall regime \cite{Geim}.
\end{abstract}
\maketitle

\textit{Introduction}. The physics of thermoelectric phenomena keeps surprising us, even though its foundations were laid almost two centuries ago \cite{Behnia_book,PED24}. In particular, this concerns the famous Seebeck effect \cite{Seebeck} that consists in the induction of a difference of potential between hot and cold edges of a conductor. The effect elucidates interconnection between electrostatic and chemical potentials that manifests itself in the stationary regime where no electric current is flowing.
The ratio of the induced voltage and applied temperature difference yields the Seebeck coefficient that constitutes an important fundamental characteristic of any conductor.
As was pointed out by Lord Kelvin in the middle of the XIXth century \cite{Kelvin}, the Seebeck coefficient can be expressed through the derivative of the chemical potential of charge carriers $\mu$ over temperature:
\begin{equation}
S_K=\frac{1}{e}\left(\frac{\partial\mu}{\partial T}\right)_{N,V},
\label{Kelvin}
\end{equation}
where $N$ is the number of particles, $V$ is the volume of the system. This relation directly follows from the constancy of the electrochemical potential
\begin{equation}
{\mu+e\phi=const},\label{constancy}
\end{equation}
with $\phi$ being the electrostatic potential characteristic of a thermalized system in a stationary state in the absence of electric current. The Kelvin's formula has proven to be almost universally valid in the broken circuit regime \cite{Shastry,KPVY2024}. Still, it has its limitations when applied to systems where non-dissipative currents may flow in the stationary regime, such as electron gases in the quantum Hall regime or topological insulators. Here we show that also in graphene, in the ballistic regime of electronic transport, the Kelvin's formula may not be exact. The failure of the Kelvin's formula in graphene manifests the realization of an anomalous Seebeck effect, where the difference of temperatures on two edges of the graphene flake induces not only the electrostatic potential drop but also it triggers two counter-propagating ballistic currents of "hot" and "cold" electrons that compensate each other exactly. In this regime, the graphene sheet plays role of the Maxwell's demon separating cold and hot particles, while, of course, this similarity becomes an illusion once the thermalisation processes in contacts are taken into account.

\textit{The Seebeck coefficient in graphene in the ballistic regime}. The Seebeck coefficient of a finite size piece of a two-dimensional conductor such as graphene can be defined as
\begin{equation}
S=\frac{\Delta V}{\Delta T}.\label{Seebeckdef}
\end{equation}
Here $\Delta T$ is the difference of temperatures of the metallic contacts located at the opposite edges of
the sample, and $\Delta V$ is the corresponding voltage induced due to the
Seebeck effect. We note that the induction of voltage ${\Delta V}$
results in the appearance of a ballistic current \cite{Ando} 
\begin{widetext}
\begin{equation}
{\cal J}_{V}=\int_{-\pi/2}^{\pi/2}\cos{\theta}d\theta\int_{-\infty}^{\infty}e\nu_{\theta}(E)v(E)\left[f\left(\frac{E-e\Delta V-\mu}{T}\right)-f\left(\frac{E-\mu}{T}\right)\right]dE,\label{current_V}
\end{equation}
\end{widetext}
with $v(E)$, $E$ being the electron
velocity and energy, $\theta$ being the carriers' propagation angle with respect to the potential gradient direction, $f$ being the Fermi-Dirac distribution. The density of  states of electrons propagating within a sector $d\theta$ is \cite{density}
\begin{equation}
\nu_{\theta}(E)d\theta= \frac{|E|}{\pi^2\hbar^{2}v_{F}^{2}}d\theta. \label{velocity-1}
\end{equation}
Here we have taken into account the combined valley and spin degeneracy factor for graphene $g=4$ \cite{degeneracy}. 

In its turn, the temperature difference $\Delta T$ applied
at the edges of the sample also will generate the current 
\begin{widetext}
\begin{equation}
{\cal J}_{T}=\int_{-\pi/2}^{\pi/2}\cos{\theta}d\theta\int_{-\infty}^{\infty}e\nu_{\theta}(E)v(E)\left[f\left(\frac{E-\mu(T+\Delta T)}{T+\Delta T}\right)-f\left(\frac{E-\mu(T)}{T}\right)\right]dE.\label{current_T}
\end{equation}
\end{widetext}
The sum of currents (\ref{current_V}) and (\ref{current_T}) must be zero, to keep the total current zero (the circuit is broken):
\begin{equation}
{{\cal J}_{V}+\cal J}_{T}=0. \label{compensation}
\end{equation}
The condition of zero total current is quite universal. It holds in any stationary system, in the broken circuit geometry. It is important to note that this zero total current condition is valid in the presence of non-local effects that make Kelvin's formula (\ref{Kelvin}) fail, as we show below.

In the limit ${\Delta V},{\Delta T}\rightarrow0$ Eq. (\ref{compensation})
takes the form
\begin{equation}
\int_{-\infty}^{\infty}\left\{\left[\left(\frac{\partial\mu}{\partial T}\right)\Delta T\!-\!e\Delta V\right]\frac{\partial f}{\partial E}\!-\!\Delta T\frac{\partial f}{\partial T}\right\}|E|dE\!=\!0.\label{compensation3-1}
\end{equation}
Using the relation $\frac{\partial f}{\partial E}=-\frac{\partial f}{\partial\mu}$ we arrive to the equality
\begin{widetext}
\begin{equation}
e\frac{\Delta V}{\Delta T}\int_{-\infty}^{\infty}\frac{|E|}{\cosh^{2}\left(\frac{E-\mu}{2T}\right)}dE=\int_{-\infty}^{\infty}\left[\left\{ \left(\frac{\partial\mu}{\partial T}\right)+\frac{E-\mu}{T}\right\} \right]\frac{|E|dE}{\cosh^{2}\left(\frac{E-\mu}{2T}\right)}.\label{compensation3-1-1}
\end{equation}
\end{widetext}


Finally, applying the definition (\ref{Seebeckdef}), one can see from (\ref{compensation3-1-1}) that the Kelvin's formula (\ref{Kelvin}) acquires the additional term $\delta S$ in the ballistic regime for 2D Dirac electrons:
	\begin{equation}
		\Tilde{S}_K=S_K+\delta S,
	\end{equation}
 with
\begin{equation}
\delta S = \frac{1}{e}\frac{{\frac{{{\pi ^2}}}{6} + 2{\rm{L}}{{\rm{i}}_2}\left( {1 + {e^{ - \frac{\mu }{T}}}}\right) - \frac{\mu }{T}\ln \left( {1+{e^{-\frac{\mu }{T}}} } \right) }}{{ \ln \left( {1+{e^{-\frac{\mu }{T}}} } \right)+ \frac{\mu }{2T}}},
\end{equation}
where ${\rm{L}}{{\rm{i}}_2}\left( z \right)$ is the polylogarithm function. 

The deviation of the Seebeck coefficient from the Kelvin's formula prediction seems surprising, at the first glance. However, it has a clear fundamental reason.
Indeed, in the ballistic regime, the local temperature cannot be defined. Between two thermalized contacts, the carriers stay out of thermal equilibrium. The distributions of carriers flowing in two directions (see Eqs. (\ref{current_V}) and  (\ref{current_T})) are described by two temperatures and two chemical potentials corresponding to the hot and cold electrodes, respectively.

\begin{figure}[h!]
\includegraphics[width=0.49\columnwidth]{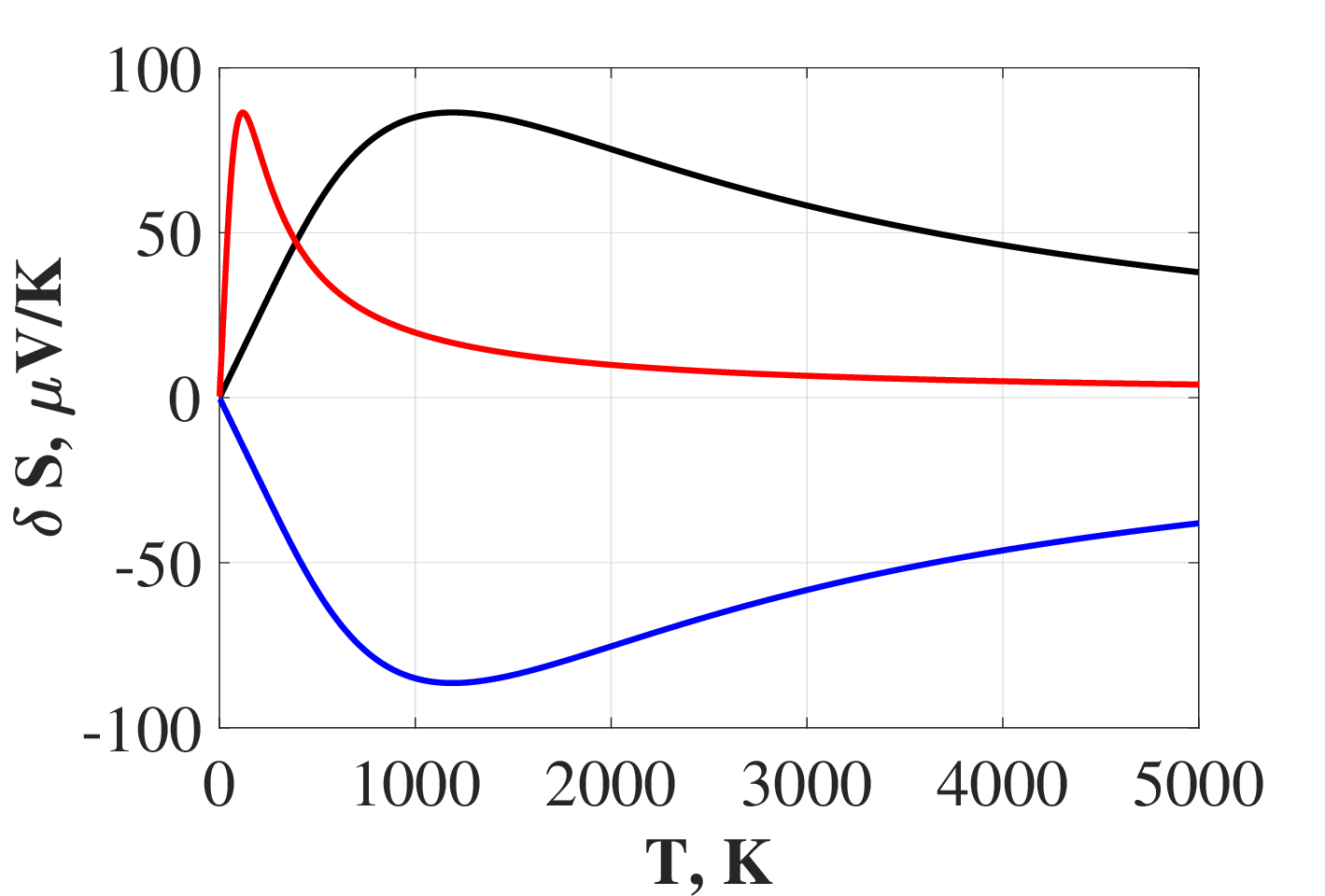}
\includegraphics[width=0.49\columnwidth]{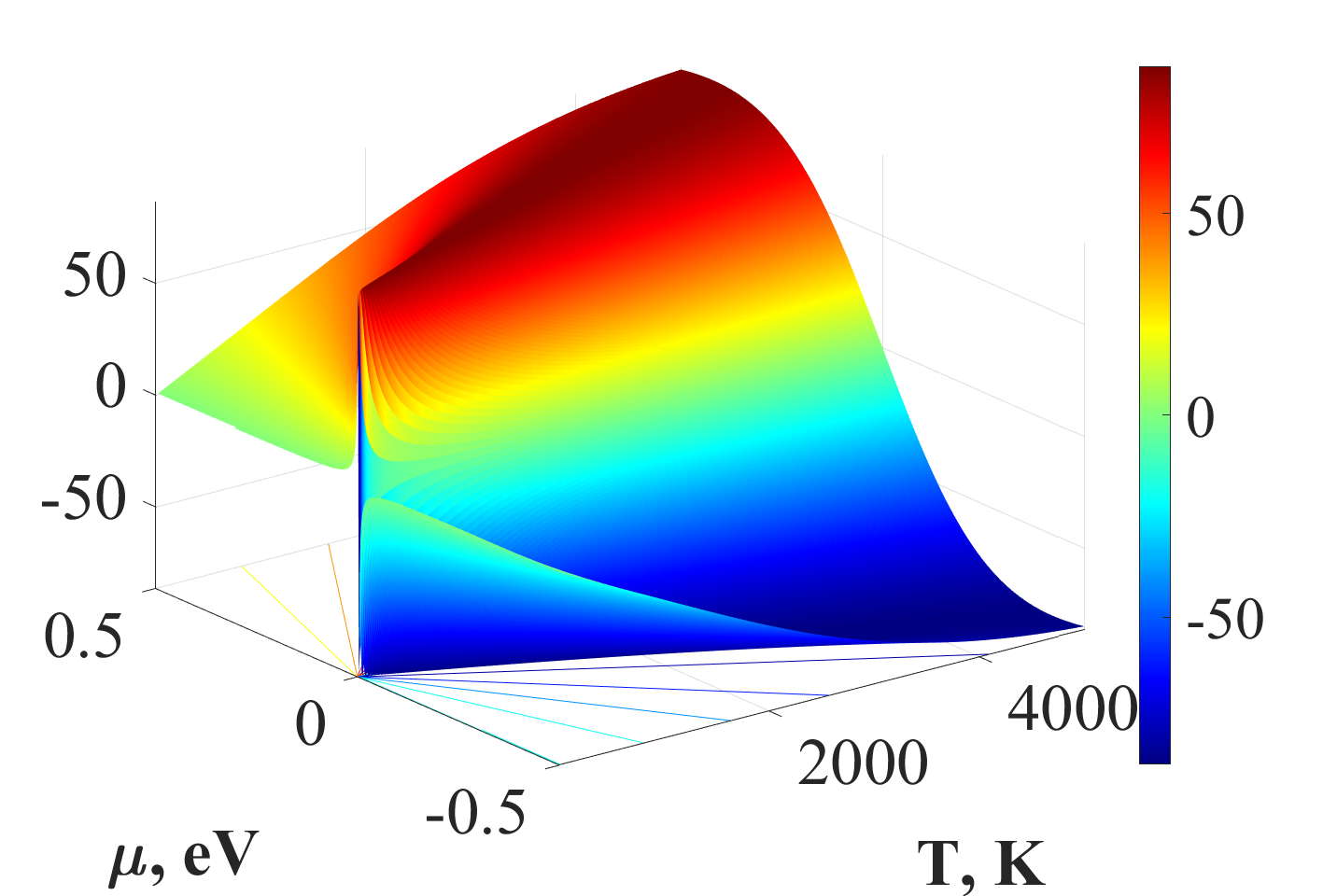}
\caption {The correction to the Seebeck coefficient (\ref{Kelvin}) arising due to the anomalous Seebeck effect $\delta S$ as a function of temperature (left panel) for the different values of chemical potential $\mu =-0.1$ eV (blue line), $\mu = 0.01$ eV  (red line), and $\mu = 0.1$ eV  (black line) and the full-scale dependence of $\delta S$ on temperature and chemical potential (right panel).}
\label{cor_Seebeck_T}
\end{figure}

Figure \ref{cor_Seebeck_T} shows the non-monotonic behavior of this correction as a function of temperature and chemical potential.  
	In the limiting cases 
	\begin{eqnarray}  \label{fin}
		\delta S = \frac{1}{e}\left\{ 
		\begin{array}{cc}
			 {\frac{{{\pi ^2}}}{3}\frac{T}{\mu } + {e^{ - \frac{\mu }{T}}} \cdot {\mathcal{O}}\left( 1 \right)}, & T \ll \mu, \\ 
	{\frac{\mu }{{T}} + {\mathcal{O}}\left( \mu^3/ T^3 \right)}, & \mu \ll T.%
		\end{array}%
		\right.   \label{finas}
	\end{eqnarray}%

 Now, in order to obtain the Seebeck coefficient of graphene explicitly in the limiting cased of low and high temperatures, we shall use the well-known expressions for the chemical potential of the 2D Dirac Fermi gas (see \cite{VK2013})
	\begin{eqnarray}  \label{f}
	\mu_D^{(2)}(T) =	\left\{ 
		\begin{array}{cc}
			\mu_D^{(2)}(0) -\frac{\pi^2 T^2}{12\mu_D^{(2)}(0)},  T \ll \mu, \\ 
		-2T\ln\left(\frac{T}{	\mu_D^{(2)}(0)}\right), \mu \ll T.%
		\end{array}%
		\right.   \label{fis}
	\end{eqnarray}
 with $\mu_D^{(2)}(0)=\hbar v_F \sqrt{2\pi n_e}$. 
  The resulting expressions for the Seebeck coefficient are 
 \begin{eqnarray}  \label{fin}
		\Tilde{S}_K =\frac{1}{e}\left\{ 
		\begin{array}{cc}
			 \frac{\pi ^2}{6}\frac{T}{\mu_D^{(2)}(0)} , & T \ll \mu_D^{(2)}(0), \\ 
	-2\ln\frac{T}{\mu_D^{(2)}(0)}, & \mu_D^{(2)}(0) \ll T.%
		\end{array}%
		\right.   \label{finas}
	\end{eqnarray}%
 
\begin{figure}[h!]
\includegraphics[width=0.49\columnwidth]{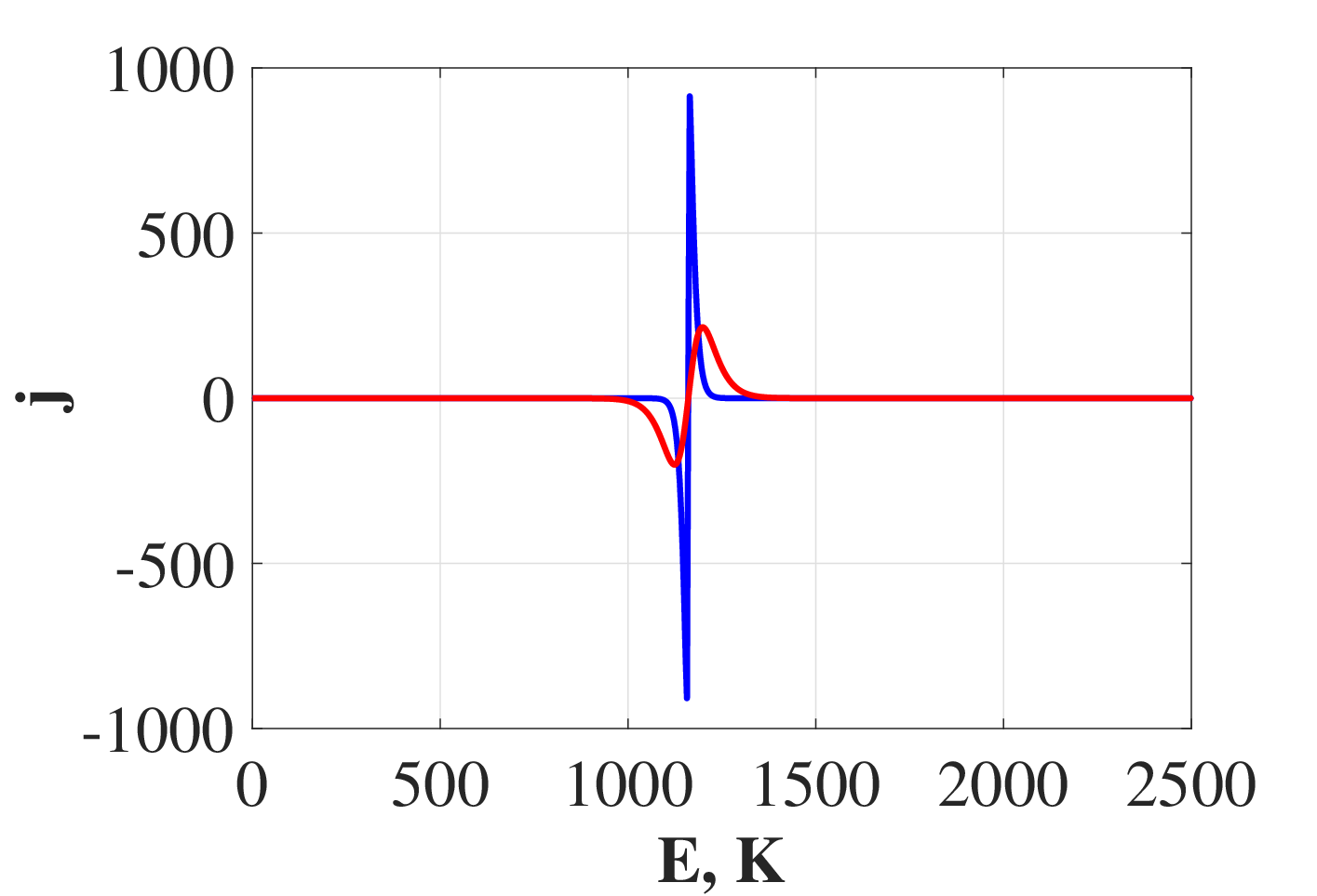}
\includegraphics[width=0.49\columnwidth]{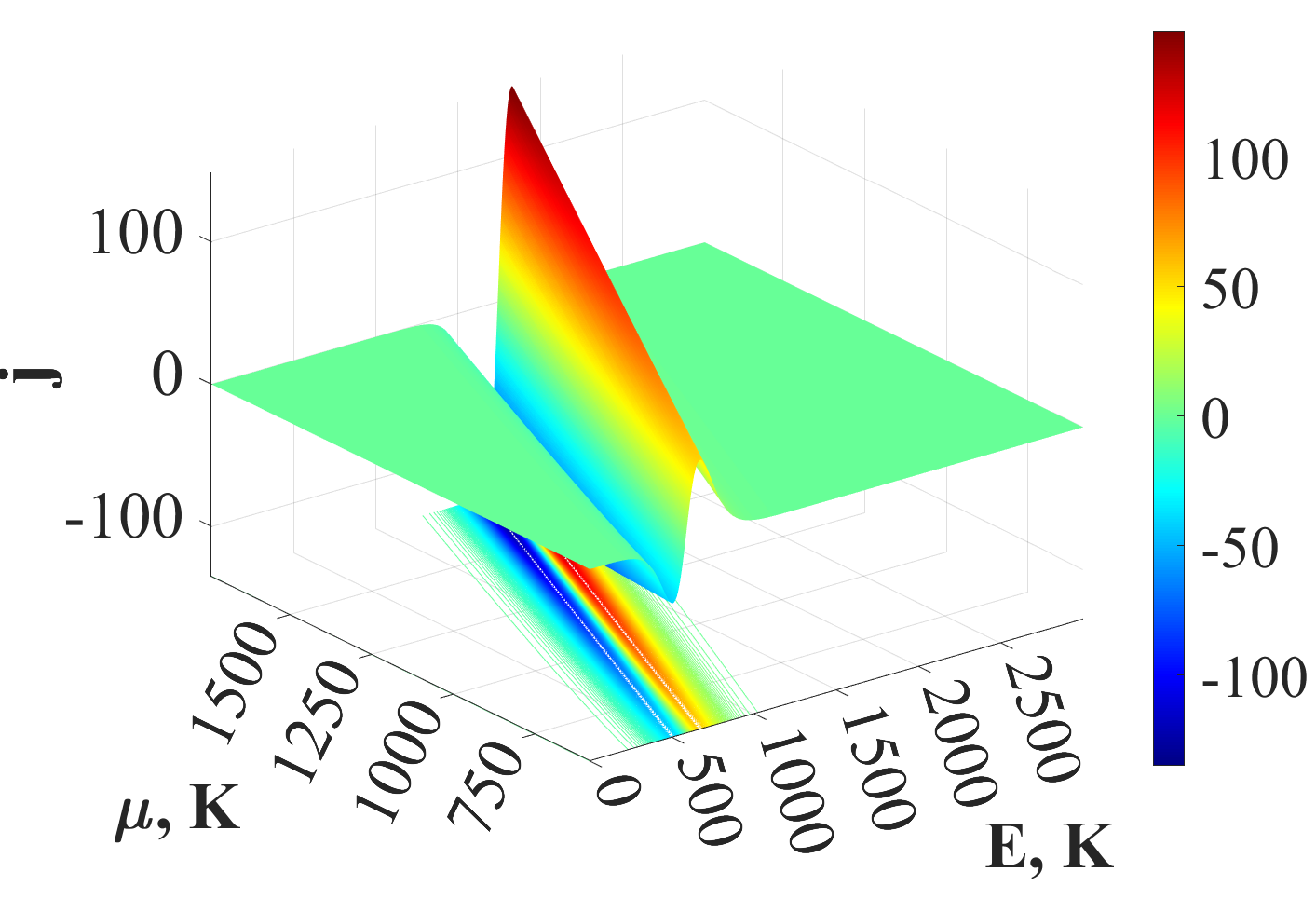}
\caption {(Left panel) The difference of two counter-propagating current densities as a function of energy for a gapless graphene with $\mu \approx 100$ meV ($\mu \approx 1160$ K) at $T=1$ K (blue line) and $T=20$ K (red line). (Right panel) The difference of counter-propagating current densities as a function of energy and chemical potential at $T=50$ K. The calculation is performed assuming $\Gamma_0=20$ K and $\Delta T = 10$ K.}
\label{curent_vs_energy}
\end{figure}

It is instructive to consider the profile of the current density distribution over energy which is determined by the integrand in Eq. (\ref{compensation})
\begin{eqnarray}
j(E)=2e\nu(E)v_{\theta}(E)\left[f\left(\frac{E-\mu(T+\Delta T)}{T+\Delta T}\right)\right.- \\ \nonumber
\left. f\left(\frac{E-e\Delta V-\mu}{T}\right)\right]. \label{current_density}
\end{eqnarray}
This function changes sign in the vicinity of the chemical potential at low temperatures. It characterises the energy dependent ballistic currents flowing through the graphene sheet in the stationary regime. Remarkably, while the total electric current is always zero in the broken circuit geometry, counter-propagating currents of carriers characterised by different average temperatures (further referred to as "hot" and "cold" carriers) may coexist compensating each other. Based on the data from \cite{VarSha}   we plot the difference of "hot" and "cold" current densities as a function of energy (see Fig. \ref{curent_vs_energy}, where we assume that $\mu(T+\Delta T) \approx \mu(T)$). One can see that the magnitude of emergent ballistic currents increases with the decrease of temperature. On the other hand, the difference of energies (or temperatures) of carriers participating in "cold" and "hot" currents decreases with the decrease of temperature. For the experimental observation of the effect, spectrally resolved current measurements would be required. Indeed, the total, spectrally integrated, current is always zero in the broken circuit geometry. The temperature of the experiment should be as low as possible but still above the limit of spectral resolution. Next, we discuss the experimental configuration that enables spatial separation of counter-propagating currents.

\textit{The asymmetry of ballistic edge currents in graphene in the quantum Hall regime.} We demonstrated above that counter-propagating currents of "hot" and "cold" electrons caused by the Seebeck effect in the ballistic regime may propagate, while the total current through the graphene sheet is zero. Now we will show how to separate these currents in real space by applying an external magnetic field. Such spatial separation would make the challenge of experimental detection of predicted manifestations of the anomalous Seebeck effect easier.
If a quantizing magnetic field is applied perpendicularly to the graphene plane,
bulk currents vanish, while ballistic edge currents propagate in the opposite directions along the opposite edges of the sample. The edge currents play a crucial role in the quantum Hall effect \cite{Hall}. If the sample is kept at a uniform temperature, the currents at the opposite edges of the Hall bar are fully symmetric \cite{GSVY2021,Altshuleretal2020}. The anomalous Seebeck effect breaks this symmetry, and it may lead to a significant imbalance of the edge currents originated from the same Landau level, but propagating in the opposite directions, as we show below.

In the presence of a strong quantizing normal-to-plane magnetic field the density of electronic states in graphene acquires the form \cite{SGB04} 
\begin{widetext}
\begin{equation}
\nu \left( {{E_n},B \to \infty } \right) = \frac{{2{\Gamma _0}}}{{{\pi ^2}{\hbar ^2}v_F^2}}\left\{ {\ln \left( {\frac{{{\Lambda ^2}}}{{2eB}}} \right) + \gamma  + \frac{{\left( {E_n^2 + \Gamma _0^2} \right)eB}}{{{{\left( {\Gamma _0^2 - E_n^2} \right)}^2} + 4E_n^2\Gamma _0^2}}} \right\},
\label{DOS_strong_magn-1}
\end{equation}
\end{widetext}
where $E_n=\sqrt{2e\hbar B v_F^2 n}$ and $\gamma$ is the Euler-Mascheroni constant,
$\Gamma_0$ is the width of Landau’s levels,
$\Lambda$ is the bandwidth (the upper limit of the integration over the energy).

\begin{figure*}
\centering
\includegraphics[width=0.49\linewidth]{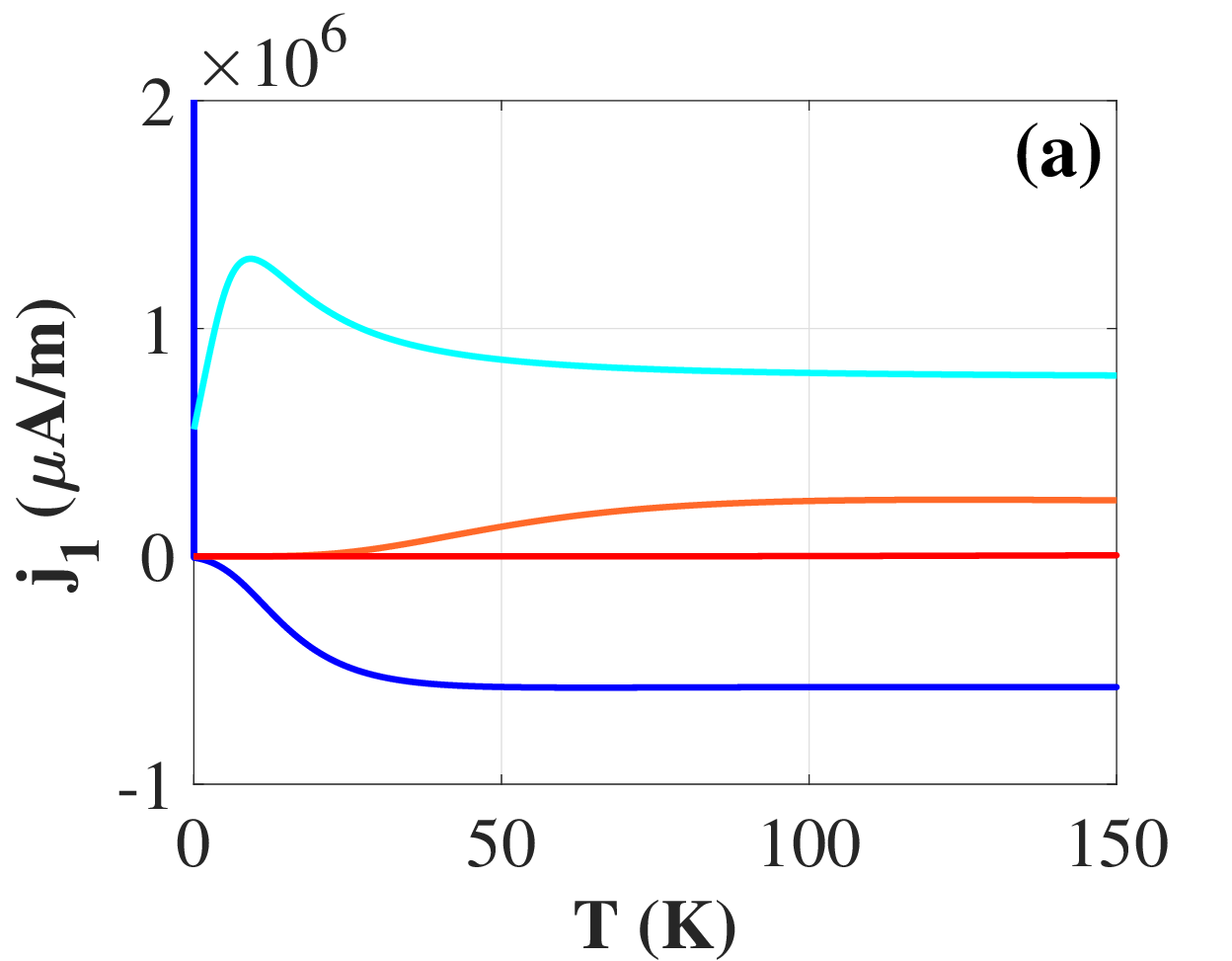}\hfil
\includegraphics[width=0.49\linewidth]{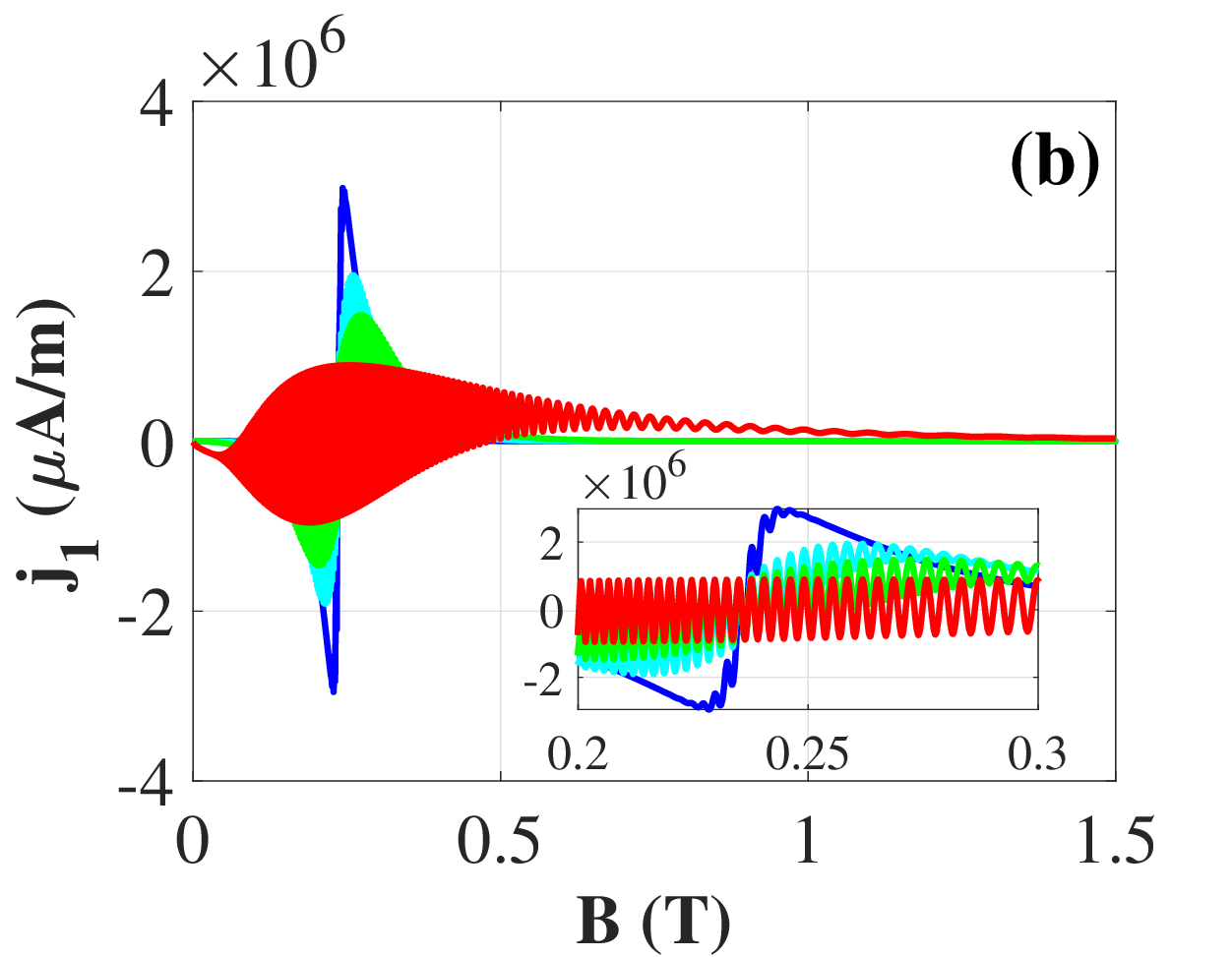}\par\medskip
\includegraphics[width=0.49\linewidth]{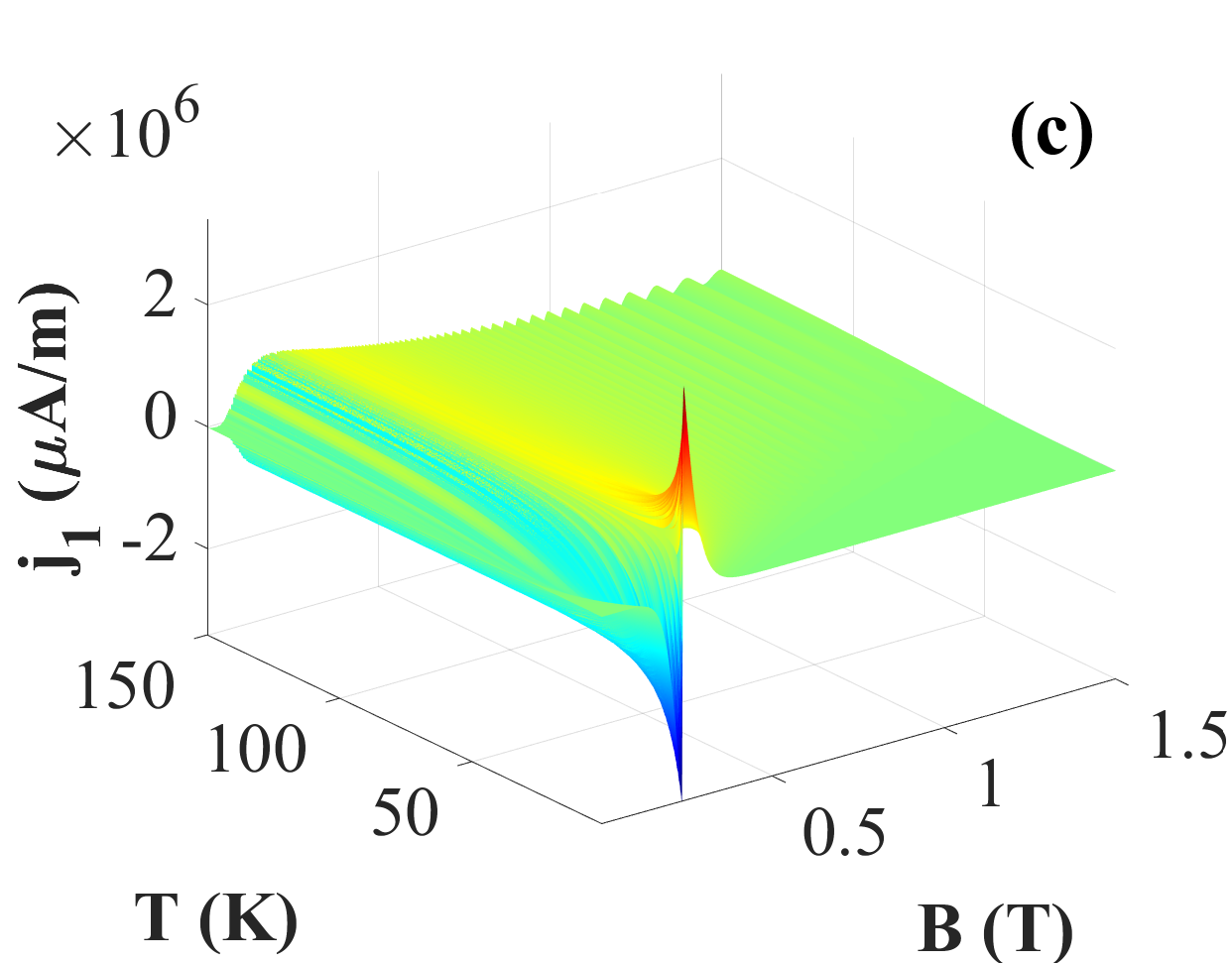}\hfil
\includegraphics[width=0.49\linewidth]{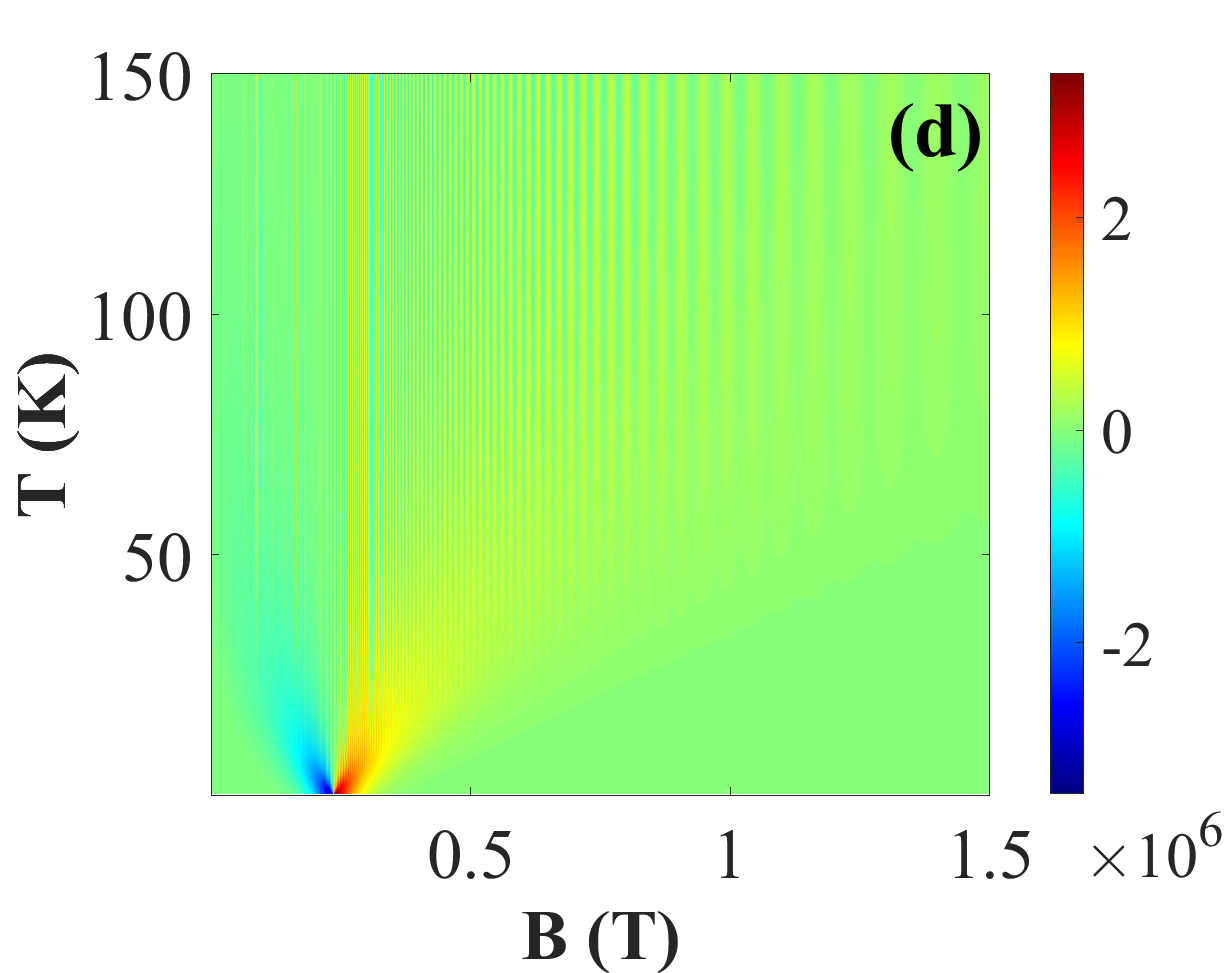}
\caption {The comprehensive behaviour of the difference of densities $j_1$ of the edge currents originated from the first Landau level $n=1$ in a gapless graphene as a function of temperature $T$ and magnetic field $B$.  (a) The temperature evolution of $j_1$ for $B=0.1$ T (blue lines), $B=0.3$ T (cyan lines), $B=1$ T (orange lines) and $B=5$ T (red lines). (b)  $j_1$ as a function of magnetic field for the fixed temperatures $T=1$ K (blue line), $T=5$ K (cyan line), $T=10$ K (green line) and $T=50$ K (red line).  The inset to panel (b) illustrates strong oscillations of $j_1$ as a function of magnetic field. The full-range dependence of the difference of edge current densities $j_1$ on temperature and magnetic field shown in the form of surface (c) and contour (d) plot. Other parameters of the calculations are $\Gamma_0=20$ K, $\Delta T = 10$ K and $\Lambda=1000$ K. }
\label{curent_vs_T_B}
\end{figure*}


We note that in the quantum Hall regime it is important to take into account the redistribution of the electron concentration between the opposite edges of the Hall bar. It is well known that the coexistence of the temperature gradient and the magnetic field leads to the stationary redistribution of the electric charge in the sample that is a manifestation of the Nernst-Ettingshausen effect (NEE). In our case, it will contribute to the asymmetry of the edge currents. Indeed, the difference of electrostatic potentials at the left and right edges of the graphene sample arising due to the NEE is given by  \cite{Lukyanchuk}:
\begin{equation}
\Delta V_{NEE}=e \nu B\Delta T.\label{Nernst}
\end{equation}
The analytical expression for the Nernst coefficient $\nu$ in graphene can be found e.g. in \cite{Lukyanchuk}. 
 Assuming that the temperature increases in $x$-direction, and the magnetic field is applied in $z$-direction, $\Delta V_{NEE}$ adds up to $\Delta V$ thus augmenting the asymmetry of two edge currents in the case of a positive Nernst constant $\nu$. In contrast, if $\nu<0$ the Nernst-Ettingshaun effect suppresses the asymmetry of edge currents. 

In order to estimate the difference of the edge currents in the limit of a strong magnetic field, let us assume that the splitting between neighboring Landau levels is much larger than temperature. In this case, one can conveniently compare the edge currents originated from the same ($n$-th) Landau level flowing along the opposite edges in opposite directions. Using Eq. (\ref{current_density}) and bearing in mind that the leading term is logarithmic, the difference of the absolute values of these two currents can be expressed as: 
\begin{widetext}
\begin{gather}
{j_n} = \frac{{2e\sqrt {2e\hbar Bn} {\Gamma _0}}}{{{\pi ^2}{\hbar ^2}}}\ln \left( {\frac{{{\Lambda ^2}}}{{2e\hbar Bv_F^2}}} \right)\left[ {f\left( {\frac{{{E_n} - {\mu _n}(T + \Delta T)}}{{T + \Delta T}}} \right) - f\left( {\frac{{{E_n} - eS\Delta T - e\nu B\Delta T - {\mu _n}(T)}}{T}} \right)} \right].
\label{current_density-1-5}
\end{gather}    
\end{widetext}

For the temperature dependence of the chemical potential, we use Eq. (17) of Ref. \cite{Lukyanchuk}.
It is important to note that the sum of $j_n$ over all Landau levels is zero, because of the broken circuit condition:
\begin{equation}
{j}=\sum\limits _{n=0}^{\infty}j_{n}=0
\end{equation}

Figure \ref{curent_vs_T_B} shows the difference of edge currents originated from the first Landau level ($n$=1) calculated as a function of temperature and magnetic field. The main peak of the current magnitude is observed at the crossing of the Fermi level and the first Landau level. The fast oscillations observed in the magnetic field dependence are due to the oscillating dependence of the chemical potential on the magnetic field. The impact of temperature on the edge currents is highly non-trivial. It is governed by the interplay of the temperature-dependent population of the first Landau level, the temperature dependencies of the chemical potential and the Nernst coefficient. The sign of the difference of two edge currents can be switched by tuning of either temperature or magnetic field.

\textit{Discussion}
The developed formalism allows for the analytical description of a non-trivial conductivity effect in broken-circuit geometries. We have demonstrated, that in the ballistic regime in a two-dimensional sample two currents of equal magnitudes flowing in the opposite directions may propagate. These currents are triggered by the Seebeck effect that leads to the induction of voltage between the hot and cold electrodes confining the sample. The total current remains zero, while the ballistic motion of electrons persists in both directions. 
One of two currents is carried by "hot" electrons (holes), the other one is carried by "cold" electrons (holes). It must be underlined that the difference between "hot" and "cold" maybe modest, but still measurable. 

As a tool for an experimental study of the predicted effect, we propose the edge current measurements in the geometry of the Nernst-Ettingshausen effect, where a quantizing magnetic field is applied perpendicularly to the plane of the sample. In this regime, the "hot" and "cold" ballistic currents are separated in real space. Each of them contributes to the edge current of the corresponding direction. 

Macroscopic ballistic edge currents in graphene bi-layers in the quantum Hall regime have been recently studied experimentally
\cite{Geim}. In the present work we consider metallic electrodes, while in Ref. \cite{Geim} superconducting electrodes are used. Also, we neglect Fabri-Perot resonances in the transmission coefficient for the charge carriers caused by the final size of the sample. If corrected to account for these two specific features of the experiment, our analytical formalism must be applicable to the description of the data \cite{Andreys_comment}. More importantly, our work predicts a new set of thermo-magnetic effects that manifest themselves in the ballistic regime. We are confident that their experimental study is feasible, and it would bring up an important new phenomenology of the energy-dependent transport of electric charge in nanostructures. 

\textit{Acknowledgement.} We thank M.E. Portnoi for critical reading of the manuscript and multiple discussions. AV is grateful to Prof. Zhanghai Chen for valuable discussions and to Xiamen University where this work was started for hospitality.


\begin{thebibliography}{1}
\bibitem{Andreys_comment} Indeed, the edge currents can be carried be Cooper pairs as it is possibly the case in Ref. \cite{Geim}. The peculiarity of pair propagation in the presence of the parallel to the edge magnetic field has been addressed in Ref. \cite{Varlamov_pair1}. In brief, the fact that Cooper pairing involves the electrons with opposite momenta means that their momenta projections on the edge direction should be opposite. At the same time the projection of momentum for a skipping electron in the Landau representation determines the position $x_0$ of the minimum of the effective potential in which the electron moves: $x_0=\lambda^2 p_y/\hbar$ (with $\lambda$ as magnetic length). Hence, one of the skipping electrons forming the pair should move along the arc corresponding to the angle $\psi \leq \pi$ with the center of rotation situated outside the sample, while the second one moves in the opposite direction along the arc corresponding to the angle $2\pi -\psi \geq \pi$, whose center of rotation is situated within the sample. In the limit of $x_0 \gg \lambda $ the pairing matrix elements for the eigen-functions belonging to so different bases are exponentially small. Hence, only the pairs that slowly move along the edge and that are characterised by the center of mass momenta less than $\hbar \lambda^{-1}$ will survive and contribute to the edge currents.
\end{thebibliography}
\end{document}